\documentstyle[12pt]{article}
\def\be {\begin{equation}}
\def\ee {\end{equation}}
\def\ba {\begin{eqnarray}}
\def\ea {\end{eqnarray}}
\newcommand{\bq}{\begin{eqnarray}}
\newcommand{\eq}{\end{eqnarray}}

\def\bi {\begin{itemize}}
\def\ei {\end{itemize}}
\begin{document}
\def\bea{\begin{eqnarray}}
\def\eea{\end{eqnarray}}
\title{\bf {Holographic Chaplygin gas model}}
 \author{M.R. Setare  \footnote{E-mail: rezakord@ipm.ir}
  \\ {Department of Science,  Payame Noor University. Bijar, Iran}}
\date{\small{}}
\maketitle
\begin{abstract}
In this paper we consider a correspondence between the holographic
dark energy density and Chaplygin gas energy density in FRW
universe. Then we reconstruct the potential and the dynamics of the
 scalar field which describe the Chaplygin cosmology.
 \end{abstract}

\newpage
\section{Introduction}
 The type Ia supernova observations
suggests that the universe is dominated by dark energy (DE) with
negative pressure which provides the dynamical mechanism of the
accelerating expansion of the universe \cite{{per},{gar},{ries}}.
The strength of this acceleration is presently matter of debate,
mainly because it depends on the theoretical model implied when
interpreting the
data.\\
An approach to the problem of DE arises from the holographic
principle that states that the number of degrees of freedom related
directly to entropy scales with the enclosing area of the system. It
was shown by 'tHooft and Susskind \cite{hologram} that effective
local quantum field theories greatly overcount degrees of freedom
because the entropy scales extensively for an effective quantum
field theory in a box of size $L$ with UV cut-off $ \Lambda$. As
pointed out by \cite{myung}, attempting to solve this problem, Cohen
{\it et al} showed \cite{cohen} that in quantum field theory, short
distance cut-off $\Lambda$ is related to long distance cut-off $L$
due to the limit set by forming a black hole. In other words the
total energy of the system with size $L$ should not exceed the mass
of the same size black hole, i.e. $L^3 \rho_{\Lambda}\leq LM_p^2$
where $\rho_{\Lambda}$ is the quantum zero-point energy density
caused by UV cut-off $\Lambda$ and $M_P$ denotes the Planck mass (
$M_p^2=1/{8\pi G})$. The largest $L$ is required to saturate this
inequality. Then its holographic energy density is given by
$\rho_{\Lambda}= 3c^2M_p^2/8\pi L^2$ in which $c$ is a free
dimensionless parameter and coefficient 3 is for convenience. As an
application of the holographic principle in cosmology,
 it was studied by \cite{KSM} that the consequence of excluding those degrees of freedom of the system
 which will never be observed by the effective field
 theory gives rise to IR cut-off $L$ at the
 future event horizon. Thus in a universe dominated by DE, the
 future event horizon will tend to a constant of the order $H^{-1}_0$, i.e. the present
 Hubble radius.
 On the basis of the cosmological state of the holographic principle, proposed by Fischler and
Susskind \cite{fischler}, a holographic model of dark Energy (HDE)
has been proposed and studied widely in the
 literature \cite{miao,HDE}.
In HDE, in order to determine the proper and well-behaved system's
IR cut-off, there are some difficulties that must be studied
carefully to get results adapted with experiments that claim our
universe has accelerated expansion. For instance, in the model
proposed by \cite{miao}, it is discussed that considering  the
particle horizon, as the IR cut-off, the HDE density reads
 \be
  \rho_{\Lambda}\propto a^{-2(1+\frac{1}{c})},
\ee
 that implies $w>-1/3$ which does not lead to an accelerated
universe. Also it is shown in \cite{easther} that for the case of
closed
universe, it violates the holographic bound.\\
The problem of taking apparent horizon (Hubble horizon) - the
outermost surface defined by the null rays which instantaneously are
not expanding, $R_A=1/H$ - as the IR cut-off in the flat universe
was discussed by Hsu \cite{Hsu}. According to Hsu's argument,
employing the Friedmann equation $\rho=3M^2_PH^2$ where $\rho$ is
the total energy density and taking $L=H^{-1}$ we will find
$\rho_m=3(1-c^2)M^2_PH^2$. Thus either $\rho_m$ or $\rho_{\Lambda}$
behave as $H^2$. So the DE results as pressureless, since
$\rho_{\Lambda}$ scales like matter energy density $\rho_m$ with the
scale factor $a$ as $a^{-3}$. Also, taking the apparent horizon as
the IR cut-off may result in a constant parameter of state $w$,
which is in contradiction with recent observations implying variable
$w$ \cite{varw}. On the other hand taking the event horizon, as
 the IR cut-off, gives results compatible with observations for a flat
 universe.\\
 In a
very interesting paper Kamenshchik, Moschella, and Pasquier
\cite{kmp}have studied a homogeneous model based on a single fluid
obeying the Chaplygin gas equation of state \be \label{chp}
P=\frac{-A}{\rho} \ee where $P$ and $\rho$ are respectively pressure
and energy density in comoving reference frame, with $\rho> 0$; $A$
is a positive constant. This equation of state has raised  a certain
interest \cite{jac} because of its many interesting and, in some
sense, intriguingly unique features. Some possible motivations for
this model from the field theory points of view are investigated in
\cite{a}. The Chaplygin gas emerges as an effective fluid associated
with d-branes \cite{b} and can also be obtained from the Born-Infeld
action \cite{c}.\\
In the present paper, we suggest  a correspondence between the
holographic dark energy scenario and the Chaplygin gas dark energy
model. We show this holographic description of the  Chaplygin gas
dark energy in FRW universe  and reconstruct the potential and the
dynamics of the  scalar field which describe the Chaplygin
cosmology.

\section{Chaplygin gas as holographic dark energy}
Here we consider the Friedmann-Robertson-Walker universe with line
element
 \be\label{metr}
ds^{2}=-dt^{2}+a^{2}(t)(\frac{dr^2}{1-kr^2}+r^2d\Omega^{2}).
 \ee
where $k$ denotes the curvature of space k=0,1,-1 for flat, closed
and open universe respectively. A closed universe with a small
positive curvature ($\Omega_k\sim 0.01$) is compatible with
observations \cite{ {wmap}, {ws}}. We use the Friedmann equation to
relate the curvature of the universe to the energy density. The
first Friedmann equation is given by
\begin{equation}
\label{2eq7} H^2+\frac{k}{a^2}=\frac{1}{3M^2_p}\Big[
 \rho_{\rm \Lambda}+\rho_{\rm m}\Big].
\end{equation}
Define as usual
\begin{equation} \label{2eq9}\Omega_{\rm
m}=\frac{\rho_{m}}{\rho_{cr}}=\frac{ \rho_{\rm
m}}{3M_p^2H^2},\hspace{1cm} \Omega_{\rm
\Lambda}=\frac{\rho_{\Lambda}}{\rho_{cr}}=\frac{ \rho_{\rm
\Lambda}}{3M^2_pH^2},\hspace{1cm}\Omega_{k}=\frac{k}{a^2H^2}
\end{equation}
Inserting the equation of state (\ref{chp}) into the relativistic
energy conservation equation, leads to a density evolving as \be
\label{enerd}\rho_{\Lambda}=\sqrt{A+\frac{B}{a^{6}}} \ee where $B$
is an
integration constant.\\
Now following \cite{bar} we assume that the origin of the dark
energy is a scalar field $\phi$, so \be \label{roph1}
\rho_{\phi}=\frac{1}{2}\dot{\phi}^{2}+V(\phi)=\sqrt{A+\frac{B}{a^{6}}}
\ee \be \label{roph2}
P_{\phi}=\frac{1}{2}\dot{\phi}^{2}-V(\phi)=\frac{-A}{\sqrt{A+\frac{B}{a^{6}}}}
\ee Then, one can easily derive the scalar potential and kinetic
energy term as \be
\label{vph}V(\phi)=\frac{2a^6(A+\frac{B}{a^{6}})-B}{2a^6\sqrt{A+\frac{B}{a^{6}}}}
\ee \be \label{phid}
\dot{\phi}^{2}=\frac{B}{a^6\sqrt{A+\frac{B}{a^{6}}}} \ee Now we
suggest a correspondence between the holographic dark energy
scenario and the  Chaplygin gas dark energy model. In non-flat
universe, our choice for holographic dark energy density is
 \be \label{holoda}
  \rho_\Lambda=3c^2M_{p}^{2}L^{-2}.
 \ee
As it was mentioned, $c$ is a positive constant in holographic model
of dark energy($c\geq1$)and the coefficient 3 is for convenient. $L$
is defined as the following form:
\begin{equation}\label{leq}
 L=ar(t),
\end{equation}
here, $a$, is scale factor and $r(t)$ is relevant to the future
event horizon of the universe. Given the fact that
\begin{eqnarray}
\int_0^{r_1}{dr\over \sqrt{1-kr^2}}&=&\frac{1}{\sqrt{|k|}}{\rm
sinn}^{-1}(\sqrt{|k|}\,r_1)\nonumber\\
&=&\left\{\begin{array}{ll}
\sin^{-1}(\sqrt{|k|}\,r_1)/\sqrt{|k|},\ \ \ \ \ \ &k=1,\\
r_1,&k=0,\\
\sinh^{-1}(\sqrt{|k|}\,r_1)/\sqrt{|k|},&k=-1,
\end{array}\right.
\end{eqnarray}
one can easily derive \be \label{leh} L=\frac{a(t) {\rm
sinn}[\sqrt{|k|}\,R_{h}(t)/a(t)]}{\sqrt{|k|}},\ee where $R_h$ is
event horizon. Therefore while $R_h$ is the radial size of the event
horizon measured in the $r$ direction, $L$ is the radius of the
event horizon measured on the sphere of the horizon. \footnote{ As I
have discussed in introduction, in non-flat case the event horizon
can not be considered as the system's IR cut-off, because if we use
$R_h$ as IR cut-off, the holographic dark energy density is given by
\be
  \rho_\Lambda=3c^2M_{p}^{2}R_{h}^{-2}.
 \ee
 When there is only dark energy and the curvature,
$\Omega_\Lambda=1+ \Omega_k$, and $c=1$, we find
 \cite{{miao2}} \be \dot R_h=\frac{1}{\sqrt{\Omega_{\Lambda}}}-1=\frac{1}{\sqrt{1+\Omega_{k}}}-1<0, \ee while we know
that in this situation we must be in de Sitter space with constant
EoS.}
 Since we have
\be\frac{\Omega_k}{\Omega_m}=a\frac{\Omega_{k0}}{\Omega_{m0}}=a\gamma,\ee
where $\gamma=\Omega_{k0}/\Omega_{m0}$, we get $\Omega_k=\Omega_m
a\gamma$ and \be \label{wmeq}
\Omega_m=\frac{1-\Omega_\Lambda}{1-a\gamma}. \ee Hence, from the
above equation, we get \be \label{wmeq1}
\frac{1}{aH}=\frac{1}{H_0}\sqrt{\frac{a(1-\Omega_\Lambda)}{\Omega_{m0}(1-a\gamma)}}.
\ee Combining Eqs. (\ref{leh}) and (\ref{wmeq1}), and using the
definition of $\Omega_\Lambda$, we obtain
\begin{eqnarray} \label{wleq} \sqrt{|k|}\frac{R_{h}}{a}&=&{\rm
sinn}^{-1}\left[c\sqrt{|\gamma|}\sqrt{\frac{a(1-\Omega_\Lambda)}{\Omega_\Lambda(1-a\gamma)}}\,\right]\nonumber\\
&=&{\rm sinn}^{-1}(c\sqrt{|\Omega_k|/\Omega_\Lambda}).
\end{eqnarray}
Using definitions
$\Omega_{\Lambda}=\frac{\rho_{\Lambda}}{\rho_{cr}}$ and
$\rho_{cr}=3M_{p}^{2}H^2$, we get

\begin{equation}\label{hl}
HL=\frac{c}{\sqrt{\Omega_{\Lambda}}}
\end{equation}
Now using Eqs.(\ref{leh}, \ref{hl}), we obtain \footnote{Now we see
that the above problem is solved when $R_h$ is replaced with $L$.
According to eqs.(\ref{2eq9}, \ref{holoda}), the ratio of the energy
density between curvature and holographic dark energy is \be
\label{ratio}
\frac{\Omega_{k}}{\Omega_{\Lambda}}=\frac{\sin^{2}y}{c^2}
 \ee
when there is only dark energy and the curvature, $\Omega_\Lambda=1+
\Omega_k$, and $c=1$, we find $\Omega_\Lambda=\frac{1}{\cos^{2}y}$,
in this case according to eq.(\ref{ldot}) $\dot L=0$, therefore, as
one expected in this de Sitter space case, the dark energy remains a
constant.
 }
\be \label{ldot}
 \dot L= \frac{c}{\sqrt{\Omega_\Lambda}}-\frac{1}{\sqrt{|k|}}\rm
cosn(\sqrt{|k|}\,R_{h}/a)
\end{equation}
where
\begin{equation}
\frac{1}{\sqrt{|k|}}{\rm cosn}(\sqrt{|k|}x)
=\left\{\begin{array}{ll}
\cos(x),\ \ \ \ \ \ &k=1,\\
1,&k=0,\\
\cosh(x),&k=-1.
\end{array}\right.
\end{equation}
By considering  the definition of holographic energy density
$\rho_{\rm \Lambda}$, and using Eqs.( \ref{hl}, \ref{ldot}) one can
find:
\begin{equation}\label{roeq}
\dot{\rho_{\Lambda}}=-2H[1-\frac{\sqrt{\Omega_\Lambda}}{c}\frac{1}{\sqrt{|k|}}\rm
cosn(\sqrt{|k|}\,R_{h}/a)]\rho_{\Lambda}
\end{equation}
Substitute this relation into following equation
\begin{eqnarray}
\label{2eq1}&& \dot{\rho}_{\rm \Lambda}+3H(1+w_{\rm
\Lambda})\rho_{\rm \Lambda} =0,
\end{eqnarray}
 we obtain
\begin{equation}\label{stateq}
w_{\rm \Lambda}=-[\frac{1}{3}+\frac{2\sqrt{\Omega_{\rm
\Lambda}}}{3c}\frac{1}{\sqrt{|k|}}\rm cosn(\sqrt{|k|}\,R_{h}/a)].
\end{equation}
If we establish the correspondence between the holographic dark
energy and Chaplygin gas energy density, then using
Eqs.(\ref{enerd},\ref{holoda}) we have \be
\label{beq}B=a^6(9c^4M_{p}^{4}L^{-4}-A) \ee Also using
Eqs.(\ref{chp},\ref{enerd}, \ref{stateq}) one can write \be
\label{aeq}w=\frac{P}{\rho}=\frac{-A}{\rho^{2}}=\frac{-A}{A+\frac{B}{a^{6}}}=-[\frac{1}{3}+\frac{2\sqrt{\Omega_{\rm
\Lambda}}}{3c}\frac{1}{\sqrt{|k|}}\rm cosn(\sqrt{|k|}\,R_{h}/a)]\ee
Substitute $B$ in the above equation, we obtain following relation
for $A$: \be
\label{aeq1}A=3c^4M_{p}^{4}L^{-4}[1+\frac{2\sqrt{\Omega_{\rm
\Lambda}}}{c}\frac{1}{\sqrt{|k|}}\rm cosn(\sqrt{|k|}\,R_{h}/a)] \ee
Then  $B$ is given by \footnote{As one can see in this case the
 $A$ and $B$ can change with time. Similar situation can arise when
 the cosmological constant has dynamic, see for example eq.(12)
 of \cite{kmp}, according to this equation
 \be A=\Lambda(\Lambda+\rho_{m}) \ee therefore, if $\Lambda$ vary
 with time \cite{shap}, $A$ does not remain constant.\\
 In the flat universe case $L$ replace with event horizon $R_h$, in
 this case equations (\ref{aeq1}, \ref{beq1})take following simple
 form respectively \be \label{aeq12}A=3c^4M_{p}^{4}R_{h}^{-4}(1+\frac{2\sqrt{\Omega_{\rm
\Lambda}}}{c})\ee
 \be \label{beq12} B=6c^4M_{p}^{4}R_{h}^{-4}
a^6(1-\frac{\sqrt{\Omega_{\rm \Lambda}}}{c}) \ee Substitute the
present value for $a$, $\Omega_{\rm \Lambda}$ and $R_{h}$, one can
obtain the values of $A$ and $B$ in present time. } \be \label{beq1}
B=6c^4M_{p}^{4}L^{-4} a^6[1-\frac{\sqrt{\Omega_{\rm
\Lambda}}}{c}\frac{1}{\sqrt{|k|}}\rm cosn(\sqrt{|k|}\,R_{h}/a)] \ee
Now we can rewritten the scalar potential and kinetic energy term as
following \bq \label{vphi1} V(\phi)=\nonumber 2c^2M_{p}^{2}L^{-2}
[1+\frac{\sqrt{\Omega_{\rm \Lambda}}}{2c}\frac{1}{\sqrt{|k|}}\rm
cosn(\sqrt{|k|}\,R_{h}/a)]\nonumber \\ =2H^2M_{p}^{2}\Omega_{\rm
\Lambda}[1+\frac{\sqrt{\Omega_{\rm
\Lambda}}}{2c}\frac{1}{\sqrt{|k|}}\rm cosn(\sqrt{|k|}\,R_{h}/a)] \eq
\be \label{phi2}
\dot{\phi}=\frac{cM_{p}}{L}\sqrt{2[1-\frac{\sqrt{\Omega_{\rm
\Lambda}}}{c}\frac{1}{\sqrt{|k|}}\rm cosn(\sqrt{|k|}\,R_{h}/a)]} \ee
Considering $x(\equiv lna)$, we have \be \label{phid}
\dot{\phi}=\phi' H \ee Then using Eqs.(\ref{hl},\ref{phi2}),
derivative of scalar field $\phi$ with respect to $x(\equiv lna)$ is
as \be \label{phi3} \phi'=M_{p}\sqrt{2\Omega_{\rm
\Lambda}[1-\frac{\sqrt{\Omega_{\rm
\Lambda}}}{c}\frac{1}{\sqrt{|k|}}\rm cosn(\sqrt{|k|}\,R_{h}/a)]} \ee
Consequently, we can easily obtain the evolutionary form of the
field \be \label{phi4}\phi(a)-\phi(a_0)=\int_{0}^{\ln
a}M_{p}\sqrt{2\Omega_{\rm \Lambda}[1-\frac{\sqrt{\Omega_{\rm
\Lambda}}}{c}\frac{1}{\sqrt{|k|}}\rm cosn(\sqrt{|k|}\,R_{h}/a)]}  dx
\ee where $a_0$ is the present time value of the scale factor.
\section{Conclusions}
 It is fair to claim that the simplicity and reasonable nature of HDE
 provide a
 more reliable framework  for investigating the problem of DE compared with other models
proposed in the literature\cite{cosmo,quint,phant}. For instance the
coincidence or "why now?" problem is easily solved in some models of
HDE based on this fundamental assumption that matter and holographic
dark energy do not conserve separately, but the matter energy
density
decays into the holographic energy density \cite{interac}.\\
Within the different candidates to play the role of the dark energy,
the Chaplygin gas, has emerged as a possible unification of dark
matter and dark energy, since its cosmological evolution is similar
to an initial dust like matter and a cosmological constant for late
times. Inspired by the fact that the Chaplygin gas possesses a
negative pressure, people \cite{mas} have undertaken the simple task
of studying a FRW cosmology of a universe filled with this type of
fluid.\\
In this paper we have associated the holographic dark energy in FRW
universe with a scalar field which describe the Chaplygin cosmology.
We have shown that the holographic dark energy can be described  by
the scalar field in a certain way. Then a correspondence between the
holographic dark energy and Chaplygin gas model of dark energy has
been established, and the potential of the holographic scalar field
and the dynamics of the field have been reconstructed.

\end{document}